# Basic Elements of Logical Graphs

Lucas Dixon

## Abstract

We considers how a particular kind of graph corresponds to multiplicative intuitionistic linear logic formula. The main feature of the graphical notation is that it absorbs certain symmetries between conjunction and implication. We look at the basic definitions and present details of an implementation in the functional programming language Standard ML. This provides a functional approach to graph traversal and demonstrates how graph isomorphism be implemented in just a few lines of readable code. This works takes the initial steps towards a graphical language and toolkit for working with logic formula and derivations.

## 1 Introduction

There is a certain redundancy in traditional accounts of logic where both conjunction and implication are treated as binary connectives. The redundancy results in a class of theorems which are interderivable and more generally a large set of statements which are equivalent. For instance, each of the following statements are interderivable:

$$(p_1 \wedge p_2) \to q \qquad (1)$$
$$(p_2 \wedge p_1) \to q \qquad (2)$$
$$p_2 \to (p_1 \to q) \qquad (3)$$
$$p_1 \to (p_2 \to q) \qquad (4)$$

These symmetries introduce problems for both interactive and automatic reasoning:

- They make it awkward to apply results. Typically, interactive proof assistants require users to re-arrange goals to allow assumptions or lemmas to be applied.

- They make it hard to lookup proved results. Looking up a lemma, axiom or assumption requires a user to state the result they are looking up in the correct form. When lookup is performed by matching with discrimination nets, a whole series of queries are needed to lookup all symmetric variants of the theorem.

- They lead to symmetries in proof search. In particular, when reasoning forward, we have to consider each derivation as distinct and thus we may search over many derivations of equivalent results.

The idea is to avoids such distinctions by treating implication and conjunction in uniform way within a restricted form of graphs, which we call *logical graphs*. This can be informally understood the following example translations between a connective-based formalism and our graphical notation:

$$
\begin{array}{c}
(p_1 \wedge p_2) \to q \\
(p_2 \wedge p_1) \to q \\
p_2 \to (p_1 \to q) \\
p_1 \to (p_2 \to q)
\end{array}
\quad \Longleftrightarrow \quad
\begin{array}{c}
\texttt{p}_1 \quad \texttt{p}_2 \\
\searrow \swarrow \\
\texttt{q}
\end{array}
\qquad
(p \to q) \to r
\quad \Longleftrightarrow \quad
\begin{array}{c}
\texttt{p} \\
\downarrow \\
\texttt{q} \\
\downarrow \\
\texttt{r}
\end{array}
$$

Edges in the graph formalism correspond to implication in the formula. To interpret a graph as the corresponding propositional formula, the graph at the source of an arrow is considered to be parenthesised more tightly than graph at the head of the arrow. Bracketing is not needed in the graphical formalism because right associated implications become multiple adjacent edges. Conversely, multiple edges entering a vertex correspond to a conjunction of premises, and multiple edges leaving a vertex are a conjunction of conclusions.

In this note, we present the graph formalism that corresponds to multiplicative intuitionistic linear logic (MILL). This very weak foundation provides a basis on which additional connectives and their structure can easily be added. We also give details of the datatypes and algorithms needed to implement the logical graphs in the functional programming language Standard ML. In particular, we develop a generic graph traversal algorithm and use this to express graph isomorphism. These developments make significant use of a library for managing sets of finite names, binary relation over these as well as mappings.

The current implementation can be found at: `http://sourceforge.net/projects/isaplanner/develop`, in the subdirectory `lgraphs`.

## 2 Multiplicative Intuitionistic Linear Logic

The *term* language for multiplicative linear logic describes proofs (for a more detailed account see [1]):

$$\texttt{t} = x \mid \lambda x.\ \texttt{t} \mid \texttt{t}_1\ \texttt{t}_2 \mid () \mid t_1 \mapsto_{()} t2 \mid (\texttt{t}_1,\ \texttt{t}_2) \mid \texttt{t}_1 \mapsto_{(x_1,\ x_2)} \texttt{t}_2$$

The *types* (also called the *formula*) of MILL are defined by:

$$\texttt{A} = \texttt{1} \mid \texttt{A}_1 \otimes \texttt{A}_2 \mid \texttt{A}_1 \multimap \texttt{A}_2$$

A *context* is a mapping of names to types which we write as a collection of pairs, separated by commas, where each pair is written $^{x:}A$ and indicates that $x$ maps to $A$. This lets us write sequents as:

$$\ldots\ ^{x:}A\ \ldots \vdash \mathtt{t} : \mathtt{A}$$

The typing rules for this language describe its logical behaviour and relate types to terms:

Rules for 1:
$$\frac{\Gamma \vdash t{:}C}{\Gamma, ^{x:}1 \vdash (x \mapsto_{()} t){:}C} \qquad \overline{\{\} \vdash ():1}$$

Rules for $\otimes$:
$$\frac{\Gamma, ^{x:}A, ^{y:}B \vdash t{:}C}{\Gamma, ^{z:}A \otimes B \vdash (z \mapsto_{(x,y)} t){:}C} \qquad \frac{\Gamma \vdash t_1{:}A \qquad \Delta \vdash t_2{:}B}{\Gamma, \Delta \vdash (t_1, t_2){:}A \otimes B}$$

Rules for $\multimap$:
$$\frac{\Gamma, ^{x:}A \vdash t{:}B}{\Gamma \vdash (\lambda x.\ t){:}A \multimap B} \qquad \frac{\Gamma \vdash t_1{:}A \qquad \Delta, ^{x:}B \vdash t_2{:}C}{\Gamma, \Delta, ^{f:}A \multimap B \vdash (t_2[f\ t_1/x]){:}C}$$

The cut rule:
$$\frac{\Gamma \vdash t_1{:}A \qquad \Delta, ^{x:}A \vdash t_2{:}B}{\Gamma, \Delta \vdash (t_2[t_1/x]){:}B}$$

Normally, the exchange rule is also explicitly stated, but here I've taken the notational convenience to represent contexts as maps. The normal freshness conditions are simply that the explicit mappings (the pairs written $^{x:}A$) are disjoint from the rest of the context (which is separated by commas).

## 3 Graphical Logic

This section formalises logical graphs which correspond to statements in MILL. These will usually be called just *graphs* for short.

**Definition 1** (Graph). *A logical graph, $G$, representing a set of MILL formula, is a pair consisting of:*

**A Vertex Labelling:** *A total surjective function $l_G : V_G \mapsto L_G$, called the labelling, which assigns labels to the vertices.*[1] *When a logical graph is drawn, labels are written within the vertices. These correspond to the propositional constants. Note that $V_G$ and $L_G$ are arbitrary sets for vertices and labels, but $V_G$ must be at least countable in size: we need an infinite source of names for occurrences of propositions. When $l(v) = x$, we will say that vertex $v$ is an instance of atom $x$, or that $v$ is labelled $x$, and write this as $v \in x$.*

---
[1] Surjectivity requires that every atom has an instance, and totality that every vertex is labelled. Surjectivity makes drawings of a logical graphs form a free algebra: the drawing captures all the information without needing to also express atoms which have no instance.

**An Edge Relation:** *A binary relation, $E_G$, over the vertices which is called the* edges *of G, or the* implication *relation. We will say that v implies w when $(v, w) \in E_G$, and we will write this as $v \to w$. Additionally, the implication relation is restricted by two further conditions:*

- Strictness: *the transitive closure of the relation forms a strict partial order. Strictness avoids self-loops and being a partial order ensures that implication is acyclic.*[2]
- Well formed:

  $$\forall x, y, z \in V_G.\ (x \to y) \land (x \to z) \iff \forall w.(y \to w) \iff (z \to w)$$

  *There is no interpretation in our graphical formalisms for graphs which cannot be parenthesised to form traditional logical statements. For example, the graphs X and Y in Figure 1, do not have a logically shaped implication relation. This condition ensures a translation to traditional formula is possible.*

  *The result is thus a restricted form of directed acyclic graph. It is sometimes useful to refer the ordering induced by the implication relation, in which case we will write $v > w$ for $v \to w$.*

We draw graphs with the vertices defined by $V_G$ and the edges by $E_G$. Within a vertex $v$ we will write its label, $l(v)$. The graphs $A$ and $B$ in Figure 1

**Proposition 1** (Graphical representation). *A logical graph is uniquely identified, up to isomorphism on the vertexes, by a graphical presentation of the graph.*

A graph is fully defined by its edges and labelling functions - this is exactly what the visual presentation captures. Furthermore, by allowing labels to be moved around a graph, we capture the idea that graphs are equivalent up to isomorphism on vertices. For instance, the following drawings are of the same graph:

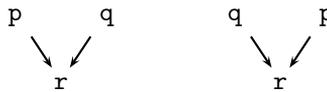

and both of these are drawings can be written as either $(l_G = \{v_0 \mapsto r, v_1 \mapsto p, v_2 \mapsto q\}, E_G = \{(v_1, v_0), (v_2, v_0)\}$ or $(l_G = \{v_0 \mapsto r, v_2 \mapsto p, v_1 \mapsto q\}, E_G = \{(v_2, v_0), (v_1, v_0)\}$. More generally, this kind of isomorphism is defined by the existence of an bijective function on the vertices that preserves the labels and their relationship to the edges. All further operations are defined up to this isomorphism of graphs, called *vertex alpha-equivalence*.

---
[2]It is also interesting to think about the general case, an arbitrary relation, where cycles are allowed. For instance, they might be used to characterise recursive algebraic datatypes. By also removing the finiteness restrictions we might also think about trying to characterise cyclic proof systems such as Brotherson's [2].

## 3.1 Implementing Graphs and Vertex Alpha-equivalence

### 3.1.1 Background Libraries

The basic code to provide a type-safe interface for using vertices and labels is:

```
structure V :> SSTR_NAMES = SStrName; (* vertices *)
structure L :> SSTR_NAMES = SStrName; (* labels *)
```

The structure `SStrName` is a type-safe interface working with names that can be made from strings (these are values of type `SStrName.name`). This structure provides an efficient implementation of finite sets of names (these have type `SStrName.NSet.T`) as well as tables from names to other objects. The naming structure also provides functions to rename sets and tables so they are disjoint.

The structures `V` and `L`, used for vertices and labels, are opaquely defined to be equal to the `SStrName`. This gives them all the functionality of `SStrName` while letting the type system distinguish between labels and vertices. This avoids any possible errors where a label is provided when a vertex is needed, or visa-versa.

The structures for the edge relations and labelling functions is:

```
structure ERel = NameBRelFun(structure Dom = V and Cod = V);
structure Labelling = NameMapFun(structure Dom = V and Cod = L);
```

In particular, the structure `ERel` provides functions for working with relations over vertices. For instance, the function `ERel.codf` takes a finite relation (an object of type `ERel.T`), and then a vertex, $v$, and returns the set of vertices that have edges going to $v$. This structure also has functions for modifying relations. For instance the function `ERel.add1` (of type `V.name -> V.name -> ERel.T -> ERel.T`) adds to a relation a new edge between the two given vertices.

The structure `Labelling` provides functions for finite maps from vertices to labels (these have type `Labelling.T`). This structure is similar to the one for finite relations, and provides essentially the same tools for modifying mappings.

Renaming machinery also comes with each of these structures. This allows the names in a relation to be made distinct from another relation, while preserving the structure. The resulting renaming is itself a mapping on finite sets of names. These can be composed and provide basis for developing renaming tools for richer data structures, such as our implementation of graphs.

### 3.1.2 Graphs

Using our libraries described above, we define labelled-graphs directly as the datatype:

```
datatype graph = Graph of
  { edges : ERel.T,
    labelling : Labelling.T
  }
```

The additional restrictions to ensure that our graphs correspond to MILL are then be enforced by using an abstract type within a signature for constructing logical graphs.

Once we have this type for graphs, we can define vertex alpha-equivalence. Computationally, this requires some work to be done. It involves identifying an isomorphism between vertices that preserves the relationship between edges and labels. This is also needed for other functions on graphs, such as checking the subgraph relation.

To develop vertex alpha-equivalence, we first define a generic and efficient functional form of graph-traversal. Then we show how it lets vertex alpha-convertibility be defined concisely and in such a way that in can be built into other kinds of graph traversal.

### 3.1.3 Generic Efficient Functional Graph Traversal

Our basic, optimised algorithm for graph traversal is a form of graph fold done depth first[3]. The signature for this is:

```
val traverse_dfs : (vertex -> 'a -> action * 'a) -> graph -> vertex -> 'a -> 'a
```

The first argument, lets call this function `f`, is applied to each vertex and the accumulated value (which has type `'a`). It results in some information on how to continue the traversal (the type `action`) as well as an updated accumulator value. The type `action` is defined by:

```
datatype action = Skip | Stop | Continue;
```

This informs the traversal algorithm of what to do next. This can be `Skip` to avoid traversal of the vertices' parents, `Stop` to exit traversal altogether and return the current accumulated value, or `Continue` to proceed traversal depth first over the vertices' parents. This integration of control flow into the folded function allows the traversal function to provide a rich set of behaviours as well as to stop early.

The generic efficient graph traversal algorithm is:

```
fun traverse_dfs f g v a0 =
   let
     val stopped_with = ref a0 (* value returned *)
     fun foldf v a =
         case f v a
          of (Skip, a2) => a2 (* skip children here *)
           | (Stop, a2) => (* stop with this value *)
             (stopped_with := a2; raise stop_exp ())
           | (Continue, a2) => (* continue with this accumulator *)
             V.NSet.fold foldf (ERel.codf (get_edges g) v) a2
   in (foldf v a0 handle stop_exp () => !stopped_with) end;
```

---

[3]This can easily be further abstracted be take the search algorithm as a parameter by using a generic notion of agenda, as described in [3].

The function `f` is the function being folded over the graph; `g` is the graph; `v` is the initial vertex from which the graph is being traversed, going backward along edges; and `a0` is the initial value of the accumulator.

The above algorithm uses imperative features of ML to stop the search early and return the intermediate value. The exception `stop_exp` is raised after setting the value of a temporary reference. Extensionally, this is equal to the less efficient pure function that continues traversal but ignores future vertices. Also, by hiding the reference value and exception locally, seen from the outside, the function is purely functional.

Note that this traversal function leaves the responsibility of terminating to the function being folded. In particular, when given a loopy graph, the `Stop` or `Skip` action needs to be used in order for traversal to terminate. For instance, to avoid repeated considerations of the same vertex, the folded function can return the `Skip` action when a vertex is seen more than once. The alpha-conversion algorithm demonstrates this.

### 3.1.4 Vertex Alpha-Equivalence for Graphs

Vertex alpha-equivalence between two graphs is implemented by finding a mapping from between the vertex names that preserves the rest of the graph's structure. This isomorphism can also be implemented as a finite map in the same way as labelling:

```
structure VMap = NameMapFun(structure Dom = V and Cod = V);
```

To construct the vertex alpha-equivalence map, it sufficed to start with two graphs, each paired with an initial vertex from which to start constructing the isomorphism. We then traverse the graphs ensuring that a subgraph isomorphism of the seen parts of the graphs is constructed. Initially there is only one possible map which is between the initial pair of vertices. As matching progresses, because there may be many possible graph isomorphisms, a list possible graph isomorphism mappings are maintained. The function to construct a graph isomorphism is then defined using graph traversal:

```
fun mk_graph_iso (g1,v1) (g2,v2) =
  traverse_dfs (iso_trav g1 g2) g1 v1 [VMap.add v1 v2 VMap.empty]
```

where the function `iso_trav` is the following auxiliary fold function that incrementally builds up the list of possible isomorphisms:

```
fun iso_trav g1 g2 v1 [] = (Stop,[])
  | iso_trav g1 g2 v1 vmaps =
  let val vasms = (get_v_asms g1 v1) in
    (Continue,
     maps (fn vmap =>
       let val v2 = (VMap.domf vmap v1)
       in vertex_match_perms (g1, vasms) (g2, get_v_asms g2 v2) vmap end)
```

```
    vmaps)
  end;
```

The location in the first graph is the vertex given to the fold function by graph traversal (`v1`). It then tries to extend each accumulated vertex mapping (members of `vmaps`) to the parent vertices. It uses the mapping of `v1` to identify the corresponding vertex in the second graph. The first case of the `iso_trav` function , when there no mappings, lets graph traversal exit early. Otherwise graph traversal will continue from vertex `v1` of graph `g1`. For each possible isomorphism (the `vmaps`), the unmapped labelling of the assumption vertices of `v1` are compared with the corresponding ones in `g2`. This is a combinatorial problem (performed by the function `vertex_match_perms`) which can result in many possible mappings ($O(n^2)$). These are the possible extensions of the existing isomorphisms. Such branching of alternative possible mappings is unavoidable - it is defined by the symmetry that is inherent in the graph. Searching over them seems unavoidable[4]. The best that can be done, beyond what is implemented above, seems to be to use lazy lists to minimise the memory overheads. This is a trivial refactoring of the above code that slightly reduces its readability.

The natural properties to prove of this algorithm are soundness and completeness. Soundness states that every resulting map is a (sub)graph isomorphism between `g1` and `g2`. Completeness states that every isomorphism containing `v1` being mapped to `v2` is returned. The crucial induction scheme, needed to prove these properties, comes from the interaction between the `vertex_match_perms` function and graph traversal. In particular, the number of unmapped vertices decreases in each mapping returned by `vertex_match_perms`, and thus . Completeness comes from the considering every possible extension to the graph isomorphisms, performed by `vertex_match_perms`, and sounders comes from each step resiting in only isomorphism mappings for the seen (sub)graph visited so far.

## 4 Basic Definitions

Given the definition of logical graphs, there are a some natural concepts from traditional accounts of logic which we would like to be able to talk about. We describe how *subgraph* corresponds to sub-term and in particular give the graphical analogue for decomposing a term into its assumptions and conclusions.

**Definition 2** (Vertex Subgraph). *$G$ is said to be a vertex subgraph of $H$, written $G \subset_v H$, when $l_G \subset l_H$ and $E_G \subset E_H$. When a set of vertices, $W$, is a subset of $V_G$ we will write $W/G$, for the subgraph of $G$ that contains precisely the vertices in $W$ as well as edges between them.*

**Definition 3** (Strict Vertex Subgraph). *$G$ is said to be a strict vertex subgraph of $H$, written $G \leq_v H$, when $G \subset_v H$ and $G$ edges between vertices exactly when $H$ does: $\forall v, w \in G. (v \to w) \in E_G \iff (v \to w) \in E_H$.*

---

[4]There is no known polynomial graph isomorphism algorithm, although the problem is also not known to be NP complete.

| A | B | X | Y |
|---|---|---|---|
| $(f \multimap g) \otimes ((a \multimap b \otimes c) \otimes d \multimap e)$ | $a \otimes b \multimap b \otimes c$ | | |

Figure 1: The graphs A and B are logical graphs hence correspond to formula in MILL, but the graphs X and Y are not logical graphs (they are not *well formed*, see Definition 1).

The concept of the *conclusions of a graph* is analogous to the conclusions of statement.

**Definition 4** (Conclusion). *We call a vertex, v, in a graph, G, a conclusion of G when it is minimal: when there are no edges from v to any other the vertex in G. We write $C_G$ for the set of vertices that are conclusions of G. For example, in Figure 1, the vertices e and g are the conclusions of the graph A.*

Similarly, we define the *assumption graph* in a way that corresponds to an assumption of a statement:

**Definition 5** (Assumption Graph). *A logical graph G is called an assumption graph of H, written $G \triangleright H$, iff $G \subset H$ and every vertex in G has every assumption it has in H. Formally, $G \triangleright H \iff G \subset H \land \forall v \in V_G. \forall w \in V_H. (w \rightarrow v) \in E_H \implies (w \rightarrow v) \in E_G$. Each vertex uniquely defines an assumption graph of which it is the conclusion. The assumption graph, within the graph H, identified by the vertex v is written $v \triangleright H$.*

By considering all assumptions of a vertex, we get its *full assumption graph*:

**Definition 6** (Full Assumption graph). *The full assumption graph of a vertex v in G, written $A(G, v)$, is the subgraph formed from the union of all assumption graphs in G that conclude in an assumption of v. Formally, $A(G, v) = \bigcup \{w \triangleright H \mid (w \rightarrow v) \in E_G\}$*

### 4.1 Implementation

Graph traversal is used for implementing the definition of assumption graph and full assumption graph. These definitions are trivial to translate directly into ML and result in linear-time functions. The subgraph operation is sensitive to the names of vertices. This allows it to be implemented in linear time by graph traversal. In particular, it does not require work modulo graph-isomorphism. This is an example of where efficiency can be gained by not providing the most general mechanism, and by instead taking advantage of the concrete names of vertices. Another example is given the following section for disjoint union of graphs.

## 5 Creating and Combining Graphs

**Definition 7** (Singleton graph). *The singleton graph for a label l is just ($\{x \mapsto l\}$, $\{\}$). This is uniquely identified by the label alone and so we can explicitly write the singleton graph of a label l by singleton(l).*

**Definition 8** (Vertex-equivalent). *Graphs that contain the same vertices and multiset of labels, but have different edges are said to be* vertex-equivalent. *Formally, two graphs, G and H, are called vertex-equivalent, written $G =_v H$, when $l_G = l_H$.*

Vertex equivalence can be used to define a graph $K$ by specifying it to be vertex-equivalent to another graph and by providing the implication relation (the edges).

**Definition 9** (Graph equivalence upto vertex-addition). *The graph G is said to be the vertex-addition of H and K when it contains the disjoint union of vertices of H and K, and has the corresponding disjoint union of labellings. We write this as $G =_v H + K$. This leaves the edges of G unspecified.*

*To define this formally, we use $X \uplus Y$ for disjoint union. The subscripts $x_1$ and $y_2$ provide indexes corresponding to left and right elements from $X$ and $Y$ respectively. We can then define $G =_v H + K$ by $V_G = V_H \uplus V_K$ such that the labelling is preserved:*

$$\forall v \in V_H.\ l_G(v_1) = l_H(v)$$
$$\forall v \in V_K.\ l_G(v_2) = l_K(v)$$

**Definition 10** (Graph Addition). *The graph G is said to be the addition of the graphs H and K, written $G = H + K$, defined by the vertex-equivalence $G =_v H + K$ and edge relation $E_G = \{(v_1, u_1) \mid (v, u) \in E_H\} \cup \{(v_2, u_2) \mid (v, u) \in E_K\}$.*

Vertex-addition can also be used to define vertex-sensitive subtraction: $G =_v H - K \iff G + K =_v H$. Graphically, this corresponds to removing part of a graph. Because labelling is surjective, any labels without instances are removed by subtraction, and thus subtraction, unlike addition, is uniquely defined. Similarly to addition, there is a natural way to define the edge relation to create the definition for *graph vertex-sensitive subtraction*:

**Definition 11** (Graph Vertex-Sensitive Subtraction). *The graph G is the result of subtracting the graph K from H, written $G = H - K$, and is defined by $G =_v H - K$ and the edge relation $E_G = \{(v_1, u_1) \mid (v, u) \in E_H \land v \notin V_K \land u \notin V_K\}$.*

Because this is sensitive to the names of vertices, it is directly not 'stable' modulo vertex alpha-equivalence with respect to addition:

$$(H + K) - H \neq K \tag{5}$$
$$G - (H + K) \neq (G - H) - K \tag{6}$$

However, if we apply the renaming performed by disjoint union of graph addition, we do get:

$$(H + K) - H_1 = K \tag{7}$$
$$(K + H) - H_2 = K \tag{8}$$

The following equations also hold modulo vertex alpha-equivalence for addition and subtraction, where 0 denotes the empty graph which forms the additive unit:

$$0 + H = H \tag{9}$$
$$G + H = H + G \tag{10}$$
$$(G + H) + K = G + (H + K) \tag{11}$$
$$H - H = 0 \tag{12}$$
$$\tag{13}$$

**Definition 12** (Graph Implies). *The graph $G$ is the called the graph of $H$ implies $K$, written $G = H \to K$, when $G =_v H + K$ and the edge relation is $E_G = E_{H+K} \cup \{v_1 \to w_2 \mid (v \in C_H) \wedge (w \in C_K)\}$ (every conclusion of $H$ has an edge going to every conclusion of $K$).*

### 5.1 Implementation

Graph addition can be implemented rather neatly using renaming and union: instead of providing a disjoint union which renames vertices in both graphs, we rename the vertices from the first graph and then simply use the union of the `Labelling` and `ERel` structures. For it's part the renaming provides a vertex mapping that avoids name clashes with the second graph. Graph implies is a simple extension of addition. Vertex sensitive subtraction (it respects vertex names) is easy to implement and linear time as it simply involves a fold on the set of vertex names. Subtraction up-to vertex alpha-equivalence requires first identifying the equivalence, then removing the image of the equivalence, this known to be NP-complete.

## 6 Translating Between Logical Graphs and MILL

Translating a MILL formula (a type) to a corresponding graph is easy to define using graph addition and recursion over the formula:

$$\llbracket A \multimap B \rrbracket_G = \llbracket A \rrbracket_G \rightarrow \llbracket B \rrbracket_G \qquad (14)$$
$$\llbracket A \otimes B \rrbracket_G = \llbracket A \rrbracket_G + \llbracket B \rrbracket_G \qquad (15)$$
$$\llbracket 1 \rrbracket_G = empty \qquad (16)$$
$$\llbracket A \rrbracket_G = singleton(A) \qquad (17)$$
$$(18)$$

The corresponding implementation looks the same.

The translation from a graph to a formula in MILL involves a special decomposition of logical graphs. This is the induction principle that arises from the acyclic and well-formedness conditions: the conclusions of a subgraph are either disjoint, or share the same set of assumptions. A graph can thus be first decomposed into disjoint parts: the full assumption graph and the conclusion vertices. Hence, every graph is either the singleton graph, or of the form:

$$G = (A(G, \bar{v_1}) \rightarrow (\bar{v_1}/G)) + ... + (A(G, \bar{v_n}) \rightarrow (\bar{v_n}/G))$$

or, without the ellipsis notation:

$$G = \Sigma_{\bar{v} \subset R_G}(A(G, \bar{v}) \rightarrow (\bar{v}/G))$$

The translation to MILL then simply defined by:

$$\llbracket singleton(l) \rrbracket_F = l \qquad (19)$$
$$\llbracket \Sigma_{\bar{v} \subset R_G}(A(G, \bar{v}) \rightarrow (\bar{v}/G)) \rrbracket_F = \otimes_{\bar{v} \subset R_G}(\llbracket A(G, \bar{v}) \rrbracket_F \multimap \llbracket \bar{v}/G \rrbracket_F) \qquad (20)$$

The equations that show that our graphical representation provides a normal form for formula are:

$$\llbracket \llbracket G \rrbracket_F \rrbracket_G = G \qquad (21)$$
$$\llbracket F \rrbracket_G \rrbracket_F \iff F \qquad (22)$$

## 6.1 Implementation

Because this decomposition of graphs depends on multiple conclusions, it cannot be directly expressed as a graph traversal. However, a direct implementation of conclusion clique can be defined without much difficulty. This identifies the clusters conclusions (the $\bar{v}$). Then the assumption graphs can be computed for a single member of a clique. This provides a destructor function.

Given the destructor function, the translation of a graph to a formula is trivial to implement.

## 6.2 Conclusions and Further Work

We have introduced the notion of logical graphs and shown a basic representation for them and given some algorithms for their implementation. The main algorithm is a general form of graph traversal. This is used to define graph isomorphism. The code is essentially functional, with imperative features used for efficiency and hidden within the traversal function. We go on to describe some further basic definitions for constructing graphs and for translating between graphs and multiplicative intuitionistic logic.

Further work includes presenting the corresponding notion of proof and proving the appropriate properties of logical graphs such as soundness and completeness with respect to MILL.